\documentclass[10pt,aps,twocolumn, prl,showpacs, footinbib,superscriptaddress]{revtex4-1}
\usepackage{graphicx}
\usepackage{psfrag}
\usepackage{amsmath,amssymb}
\usepackage{colordvi}
\usepackage{color}

\renewcommand{\a}{\alpha}
\renewcommand{\b}{\beta}
\newcommand{\g}{\gamma}

\newcommand{\s}{\sigma}
\newcommand{\D}{\Delta}
\renewcommand{\th}{\theta}

\newcommand{\e}{\epsilon}
\renewcommand{\l}{\lambda}

\newcommand{\ua}{\uparrow}
\newcommand{\da}{\downarrow}

\renewcommand{\dag}{\dagger}

\newcommand{\nn}{\nonumber}

\newcommand{\lr}[1]{\left\langle#1\right\rangle}

\newcommand{\be}{\begin{equation}}
\newcommand{\ee}{\end{equation}}
\newcommand{\ba}{\begin{eqnarray}}
\newcommand{\ea}{\end{eqnarray}}
\newcommand{\m}[4]{\left[\begin{array}{cc} #1&#2\\#3&#4 \end{array}\right]}

\renewcommand{\v}[2]{\left[\begin{array}{c} #1\\#2 \end{array}\right]}

\begin{document}
\title{Mott-Superfluid Transition for Spin-Orbit Coupled Bosons in One-Dimensional Optical Lattices}
\author{Zhihao Xu}
\affiliation{Department of Physics and Center of Theoretical and Computational Physics, The University of Hong Kong, Hong Kong, China}
\author{William S. Cole}
\affiliation{Department of Physics, The Ohio State University, Columbus, Ohio 43210, USA}
\author{Shizhong Zhang}
\affiliation{Department of Physics and Center of Theoretical and Computational Physics, The University of Hong Kong, Hong Kong, China}
\date{\today}
\begin{abstract}
We study the effects of spin-orbit coupling on the Mott-superfluid transition of bosons in a one-dimensional optical lattice. We determine the strong-coupling magnetic phase diagram by a combination of exact analytic and numerical means. Smooth evolution of the magnetic structure into the superfluid phases is investigated with the density matrix renormalization group technique. Other magnetic phases are seen and phase transitions between them within the superfluid regime are discussed. Possible experimental detection is discussed.
\end{abstract}

\maketitle

{\em Introduction}. Recent progress in cold atom physics has made it possible to use Raman lasers to generate a synthetic spin-orbit coupling for both bosonic~\cite{Spielman09L,Spielman09N,Spielman11N,ShuaiChen12L,Spielman2013N,Ji2014} and fermionic~\cite{Cheuk2012,JingZhang12L,JingZhang13A,JingZhang14N} atoms. When combined with optical lattices, it has been pointed out that in addition to the standard spin-conserving tunneling between nearest-neighbor lattice sites, an additional spin-flip term is generated. In the limit of strong lattice confinement this leads to the so-called Dzyaloshinskii-Moriya (DM) exchange interaction, as well as anisotropic couplings~\cite{Cole2012,Radic12,ZiCai12A}. In those works it was shown that these terms generate a wealth of magnetic phases for the case of two-dimensional optical lattices. Other work has emphasized the role that magnetic order (or phase order) plays in the Mott-superfluid transition~\cite{Cole2012,Grass2011,QianScarola2013}.

So far, however, only one-dimensional (``Rashba + Dresselhaus") spin-orbit coupling has been realized in experiments~\cite{Spielman09L,Spielman09N,Spielman11N,Ji2014, Cheuk2012,ShuaiChen12L,Spielman2013N,JingZhang12L,JingZhang13A,JingZhang14N} and it is therefore of interest to investigate the corresponding limit of the model studied in Refs.~\cite{Cole2012,Radic12,ZiCai12A}. Additionally, by restricting our attention to one spatial dimension we can go beyond the mean-field results described in those works. In particular, the question of how the magnetic order from the strong-coupling limit evolves through the Mott-superfluid transition can be studied with the numerically exact density matrix renormalization group (DMRG) method~\cite{White,Schollwock}.
%For example, in a certain parameter regime, a weak coupling analysis predicts the emergence of a chiral superfluid in which the phase of the superfluid order parameter modulates along the one-dimensional chain, reminiscent of the chiral magnetic phase that appears in the strong coupling limit.

The one-dimensional Hamitonian is of the following form~\cite{Cole2012}:
\begin{align}
H =-t\sum_{\lr{ij}}(\psi^\dag_i\mathcal{R}_{ij}\psi_j+\mbox{H.c.})+\frac{1}{2}\sum_{i\b\b'}U_{\b\b'}a^\dag_{i\b}a^\dag_{i\b'}a_{i\b'}a_{i\b},
\end{align}
where $\psi^\dag_i=(a^\dag_{i\ua},a^\dag_{i\da})$ and $a^\dag_{i\b}$ creates a boson with spin $\b=\ua,\da$ at site $i$. The hopping matrix has the following form $\mathcal{R}_{ij}=\cos\a\pm i\sin\a\s_y$, along the $\pm\hat{x}$ direction, respectively. The diagonal terms of $\mathcal{R}$ describe spin-conserving hopping while its off-diagonal terms describe spin-flipping hopping between nearest neighbors. To simplify the discussion, we set $U_{\ua\ua}=U_{\da\da}\equiv U$ and $U_{\ua\da}=U_{\da\ua}\equiv\lambda U$, as in Ref.\cite{Cole2012}. 
The parameter $\alpha$ is determined by the ratio of the wave vector $k_{\rm soc}$ describing the momentum transfer from the Raman lasers to the wave vector of the optical lattice $k_{\rm ol}$, $\alpha = \pi (k_{\rm soc} / k_{\rm ol})$~\cite{Radic12,ZiCai12A}.

{\em Magnetic phases in the Mott insulator}. In the limit when both $U,\lambda U\gg t$, one can perform second order perturbation theory and obtain an effective spin Hamiltonian describing the low energy dynamics in the Mott insulating states. From the original boson degrees of freedom, we write local spin operators at site $i$ as ($\hbar=1$) $\vec{s}_i=\frac{1}{2}a_{i\a}^\dag\vec{\s}_{\a\b}a_{i\b}$, where $\vec{\s}=(\s_x,\s_y,\s_z)$ are the Pauli matrices. In the case of a one-dimensional optical lattice, the explicit form of the effective spin Hamiltonian is given by~\cite{Cole2012} (after rotating around the $\hat{x}$ axis by an angle $\frac{\pi}{2}$)
\begin{align}\nn
\label{magH}
H_{\rm mag} &= -\frac{4t^2}{U}\sum_{\lr{ij}}\Big[\frac{\cos(2\a)}{\l}s^x_is^x_j+\frac{\cos(2\a)}{\l}(2\l-1)s^y_is^y_j \\
&+\frac{1}{\l}s^z_is^z_j+\sin(2\a)(s^x_is^y_j-s^y_is^x_j)\Big].
\end{align}
The term proportional to $\sin(2\a)$ is the so-called Dzyaloshinskii-Moriya term~\cite{Dzyaloshinskii,Moriya} and can be written generally as $\vec{D}\cdot(\vec{s}_i\times\vec{s}_j)$, with a DM vector $\vec{D}=\sin(2\a)\hat{z}$. A similar Hamiltonian with the DM term has been proposed to describe certain quasi-one-dimensional antiferromagnets; for example, copper benzoate~\cite{Dender1996}, as suggested originally in Ref.~\cite{Oshikawa1997}. In these materials, the ratio of the possible DM interaction to the exchange interaction energy scale is very small (see Ref.~\cite{Shiba2000}, and references therein). In $H_{\rm mag}$, this ratio can be tuned arbitrarily by changing the parameter $\a$ which is easily implemented by adjusting the intensity or polarization of the Raman lasers~\cite{Spielman09L,Spielman09N,Spielman11N,Ji2014, Cheuk2012,ShuaiChen12L,Spielman2013N,JingZhang12L,JingZhang13A,JingZhang14N}.

The general spin Hamiltonian $H_{\rm mag}$ for arbitrary $\a$ and $\l$ cannot be solved exactly. In the following, we discuss a few special cases where the magnetic Hamiltonian equation~(\ref{magH}) can be dealt with analytically. From these limits, we map out of the phase diagram of the magnetic Hamiltonian, as well as the elementary excitations around stable phases. We also confirm the phase diagram with DMRG calculations. For symmetry reasons, we only have to consider $0<\a<\frac{\pi}{2}$. For $\a=0$ (no DM interactions), $H_{\rm mag}$ is the standard $XXZ$ model. In this limit, when $\l<1$, $H_{\rm mag}$ has a paramagnetic ground state, while for $\l>1$, it describes a ferromagnet along the $\hat{y}$ direction. We now address three limits that can be solved even for $\a \neq 0$.

(1) When $\l=1$, $H_{\rm mag}$ reduces to
\begin{align}\nn
H_{\rm mag} &= -\cos(2\a)\frac{4t^2}{U}\sum_{\lr{ij}}\Big[s^x_is^x_j+s^y_is^y_j  + \frac{1}{\cos(2\a)}s^z_is^z_j \\
&+ \tan(2\a)(s^x_is^y_j-s^y_is^x_j)\Big].
\end{align}
First we make the following transformation: $\tilde{s}_i^+ \equiv\exp(-i\th_i s_z)s_i^+ \exp(i\th_i s_z) = \exp(-i\th_i)s_i^+$, where $s_i^+=s_x+is_y$ is the spin-raising operator, while $\tilde{s}_i^z=s_i^z$\cite{Perk}. Choosing $\th_{i+1}-\th_i=-2\a$, the Hamiltonian reduces to an isotropic ferromagnetic Heisenberg model in terms of the $\tilde{s}_i$ spins for any $\a$. That is, $H^{(ij)}_{\rm mag}=-\frac{4t^2}{U}[\tilde{s}^x_i\tilde{s}^x_j+\tilde{s}^y_i\tilde{s}^y_j+\tilde{s}^z_i\tilde{s}^z_j]$. The ground state is a ferromagnet and the elementary excitations are spin waves with quadratic dispersion. This translates, for the original spins, to a spiral state with wave vector $2\a$ along the chain.

\begin{center}
\begin{figure}[t]
\includegraphics[width=0.5 \textwidth]{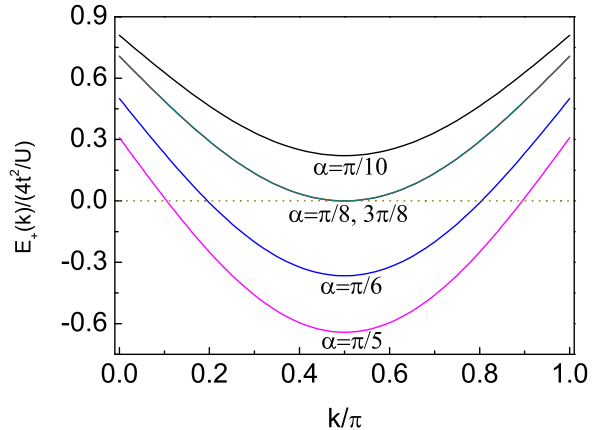}
\caption{(Color online) Excitation spectrum $E_{+}(k)$ of $H_{\rm fermion}$ in the limit $\l\to\infty$. For $\a=\frac{1}{8}\pi,\frac{3}{8}\pi$, the spectrum becomes gapless and corresponds to the transition to the gapless $xy$-chiral state. For $\a<\frac{1}{8}\pi$, the system is a $\hat{y}$ ferromagnet and for $\a>\frac{3}{8}\pi$, a $\hat{y}$ antiferromagnet; both are gapped.}
\label{fig1}
\end{figure}
\end{center}

(2) When $\l\to \infty$, $H_{\rm mag}$ takes a particularly simple form: $H^{(ij)}_{\rm mag}=-\frac{4t^2}{U}[2\cos(2\a)s^y_is^y_j+\sin(2\a)(s^x_is^y_j-s^y_is^x_j)]$. While this Hamiltonian cannot be transformed to another solved spin model, it can be solved by introducing Jordan-Wigner (JW) fermions~\cite{Sachdev} $c_i$ and $c_i^\dag$: $s_i^x=\frac{1}{2}\prod_{j<i}(1-2c_j^\dag c_j)(c_i+c_i^\dag)$ and $s_i^y=\frac{1}{2i}\prod_{j<i}(1-2c_j^\dag c_j)(c_i-c_i^\dag)$. The magnetic Hamiltonian transforms into a free fermion Hamiltonian $H_{\rm fermion}$ that can be conveniently written in the Nambu form as
\be\label{Hfermion}
H_{\rm fermion}= \sum_{k>0} [c_k^\dag,c_{-k}]\m{\e(k)}{\D(k)}{\D^*(k)}{-\e(-k)}\v{c_k}{c_{-k}^\dag},
\ee
where $\D(k)=i\cos(2\a)\sin k$ and $\e(k)=-\cos(k-2\a)$. In terms of $\Psi^\dag_k\equiv [c_k^\dag,c_{-k}]$, $H_{\rm fermion}=\sum_{k>0}\Psi^\dag_k \hat{\mathcal{H}}(k)\Psi_k$, with $\hat{\mathcal{H}}(k)=\mathcal{H}_0(k)\hat{I}+\sum_{i=x,y,z}\mathcal{H}_i(k)\hat{\s}_i$, where $\hat{I}$ is the 2-by-2 identity matrix and $\hat{\sigma}_{x,y,z}$ are the Pauli matrices. $\mathcal{H}_0(k)=-\sin2\a\sin k$, $\mathcal{H}_x(k)=0$, $\mathcal{H}_y(k)=-\cos 2\a\sin k$ and $\mathcal{H}_z(k)=-\cos 2\a\cos k$. The spectrum of fermion modes is given by $E_\pm(k)=-\sin(2\a)\sin k\pm |\cos(2\a)|$. As shown in Fig.~\ref{fig1}, the critical values for $\a$ where the spectrum $E_\pm(k)$ becomes gapless are given by $\a=\frac{1}{8}\pi,\frac{3}{8}\pi$ and this is confirmed by the DMRG calculations. For $\a<\frac{1}{8}\pi$, the system is a $\hat{y}$ ferromagnet, while for $\a>\frac{3}{8}\pi$, it is a $\hat{y}$-anti-ferromagnet. Both phases are gapped. In the intermediate region, $\frac{1}{8}\pi<\a<\frac{3}{8}\pi$, it is in the $xy$-chiral phase with gapless excitations.

\begin{center}
\begin{figure}[t]
\includegraphics[width=0.5 \textwidth]{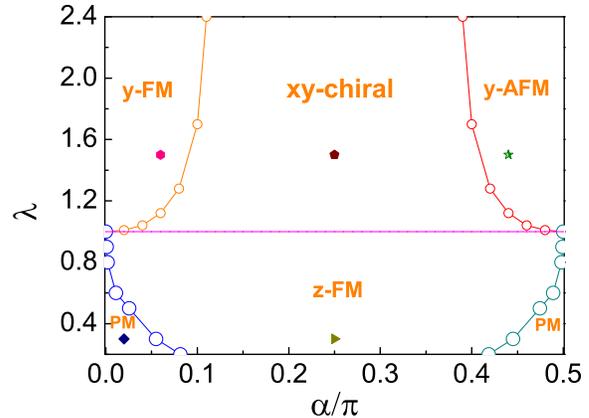}
\caption{(Color online) Phase diagram of the effective spin model $H_{\rm mag}$ in the $\lambda$-$\alpha$ plane. Five different phases are obtained. For $\l<1$, one obtains paramagnetic phases (PM) and a ferromagnetic phase along the $\hat{z}$ direction ($\hat{z}$-FM). For $\l>1$, there are three phases: a ferromagnet along the $\hat{y}$ direction ($\hat{y}$-FM), an antiferromagnet along the $\hat{y}$-direction ($\hat{y}$-AFM), and the $xy$-chiral phase. The phase boundary between $\hat{z}$-FM and $xy$ chiral are given by the straight line $\l=1$. A few representative points are marked on the phase diagram and their corresponding correlation functions are presented in Figs.~\ref{fig3} and \ref{fig4}.}
\label{fig2}
\end{figure}
\end{center}

The Hamiltonian equation~(\ref{Hfermion}) describes $p$-wave pairing in one dimension, analogous to the Kitaev model~\cite{Kitaev2001}. The Hamiltonian obeys the following symmetry: $\hat{\mathcal{H}}(k)=-\s_x\hat{\mathcal{H}}(-k)^*\s_x$ and belongs to the ``{\it D}" symmetry class, characterized by a $\mathbb{Z}_2$ invariant~\cite{Ryu2010}. In the special case when $\a=0,\frac{\pi}{2}$, it reduces to the standard Kitaev model. What is particularly interesting in this case is that the gap-closing transition for the JW fermions in the limit $\l\to\infty$ faithfully describes the finite $\l>1$ magnetic transitions shown in Fig.~\ref{fig2}.

(3) When $\a=\frac{\pi}{4}$, $H^{(ij)}_{\rm mag}=-\frac{4t^2}{U}[\frac{1}{\l}s^z_is^z_j+(s^x_is^y_j-s^y_is^x_j)]$. This is a one-dimensional Ising model with DM interactions and has been studied in the literature~\cite{Jafari2008}. It has two phases: for $\l>1$, the DM term dominates and the system is in a chiral phase in which the spin spirals around the $\hat{z}$ axis along the chain. We refer to this as the chiral $xy$ magnet. For $\l<1$, the ferromagnetic term dominates and the system is in a ferromagnetic state, pointing along the $\hat{z}$ direction. The ferromagnet has the usual Ising twofold ground state degeneracy.

%-----------------------figure--------------------------------------------------
%-----------------------figure--------------------------------------------------

The full strong-coupling magnetic phase diagram is presented in Fig.~\ref{fig2}. To distinguish between different phases, we have made use of the following set of order parameters: the gaps $\Delta_n=E_n-E_0$, measuring the energy gap between the $n$th excited state and the ground state~\cite{JiZeZhao}; the spin-spin correlation function $\mathcal{S}^\g(i,j)\equiv \langle s_i^{\gamma}s_j^{\gamma}\rangle$; the chiral correlation function $\mathcal{A}^\g(i,j)\equiv\langle A_i^{\gamma}A_j^{\gamma}\rangle$, where $\gamma=x,y,z$ and $A_i^{\gamma}=\varepsilon^{\gamma \mu \nu}(s_i^{\mu}s_{i+1}^{\nu}-s_i^{\nu}s_{i+1}^{\mu})$, describing the chirality of the spins in the ground state. Furthermore, we have calculated the entanglement entropy $S_{\rm E}=-{\rm tr}\rho_{\rm A}\ln \rho_{\rm A}$~\cite{Vedral08}, where $\rho_{\rm A}$ is the reduced density matrix, corresponding to half of the chain. At the transition point, we expect $S_{\rm E}$ to be maximal~\cite{Vedral08,Fazio,Nielsen,Kitaev,Richter,ShijiangGu}. 

\begin{center}
\begin{figure}[t]
\includegraphics[width=0.5 \textwidth]{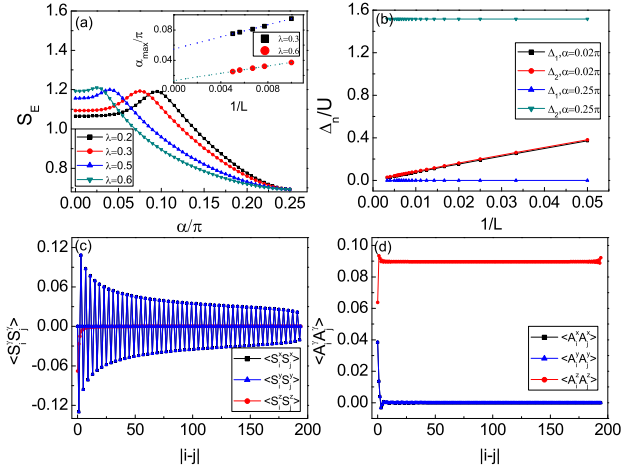}
\caption{(Color online) (a) Entanglement entropy $S_{\rm E}$ as a function of $\a$ for various values of $\l$, calculated with system size $L=200$ and open boundary condition. The peak positions $\a_{\rm max}$ determine the phase transition points. The inset shows the finite size scaling of $\a_{\rm max}$. (b)  Gaps $\Delta_n\equiv E_n-E_0$ vs $1/L$ for the representative points
of the PM ($\a=0.02\pi,\l=0.3$) and $\hat{y}$-FM ($\a=0.25\pi,\l=0.3$) regimes. (c) Spin-spin correlation function $\mathcal{S}^\g(i,j)$ in the $xy$-chiral phase. Note that $\mathcal{S}^{x,y}(i,j)$ oscillates due to the chiral nature of the phases. (d) Chiral correlations $\mathcal{A}^\g(i,j)$ in the $xy$-chiral phase. The only non-zero component is $\mathcal{A}^z(i,j)$. Both (c) and (d) are calculated with $\a=0.25\pi,\l=1.5, L=200$ and open boundary condition.}
\label{fig3}
\end{figure}
\end{center}

There are five different magnetic phases obtained within the DMRG calculations~\cite{com1}. For $\l<1$, we determine the phase boundary between the paramagnetic (PM) state and the ferromagnetic state along $\hat{z}$ axis ($\hat{z}$-FM) by locating the maximal values of $S_E$ in the $\l$-$\a$ plane, as shown in Fig.~\ref{fig3}(a), where we plot $S_{\rm E}$ as a function of $\a$ for various values of $\l$. In the inset, the value of $\a_{\rm max}$ corresponding to the maximal $S_{\rm E}$ is plotted for fixed values of $\l$ and for different system sizes and extrapolation to infinite system size is taken to identify the transition point. In Fig.\ref{fig3}(b), we have also calculated the gap for various values of $\a$ ($\l=0.3$). The phase boundary obtained from the vanishing of $\Delta_2$ is consistent with that from the maximal entanglement entropy. For $\l>1$, one finds three different phases: ferromagnetic along $\hat{y}$ ($\hat{y}$-FM), antiferromagnetic along $\hat{y}$ ($\hat{y}$-AFM), and the $xy$-chiral state in between. We calculate the chiral correlator $\mathcal{A}^\g(i,j)$, $\g=x,y,z$ and define the asymptotic value $\mathcal{A}^\g\equiv \lim_{|i-j|\to\infty}\mathcal{A}^\g(i,j)$. The $xy$-chiral phase is characterized by a non-zero value of $\mathcal{A}^z$. As an example, we show in Fig.~\ref{fig3}(c) and \ref{fig3}(d) how the spin-spin correlation function $\mathcal{S}^\g(i,j)$ and chiral correlation function $\mathcal{A}^\g(i,j)$ depend on $|i-j|$ for $\a=\frac{1}{4}\pi$ and $\l=1.5$. We note that $\mathcal{S}^x(i,j)=\mathcal{S}^y(i,j)$ oscillates and their envelope function decays algebraically. For $\mathcal{A}^\g(i,j)$, we find only one of its components, $\mathcal{A}^z(i,j)$, is non-zero and remains a constant in the chiral states. We finally note that the phase boundary between the $\hat{z}$-FM and $xy$-chiral phase is the straight line at $\l=1$ as shown before.

%-----------------------figure--------------------------------------------------
%-----------------------figure--------------------------------------------------

%-----------------------figure--------------------------------------------------
%-----------------------figure--------------------------------------------------

{\em Mott-superfluid transition}. We now discuss how the magnetic ordering in the Mott insulating state evolves into the superfluid state as one increases the hopping amplitude $t$. A similar question has been discussed in the two-dimensional case~\cite{Cole2012} with the conclusion that there is a smooth evolution of the magnetic correlations across the Mott-superfluid transition within mean-field theory (MFT). However, the question of how they match the magnetic structure in the weak-coupling superfluid phases is left undiscussed, due to limited applicability of MFT. In the one-dimensional case considered here, the situation is similar but can be studied essentially exactly. To characterize the superfluid state, we make use of the same set of correlation functions $\mathcal{S}^\g(i,j)$ and $\mathcal{A}^\g(i,j)$ defined for the strong-coupling regime, but now written in terms of the original boson operators. 

To gain insight, let us first consider the weak-coupling limit, $U\to 0$. There are two degenerate minima in the single-particle spectrum, located at $k=\pm \a$. The corresponding spinor wave functions are equal superpositions of spin-up and spin-down components, $\Psi_\pm(x)=\exp(\pm i\a x)(1,\pm i)$ at $k=\pm \a$, respectively. There are only two types of superfluid states, corresponding to different magnetic structures in the weak coupling limit~\cite{Huhui,Galitski,Shizhong,Zhaihui10L,Baym,CZhang,LiYun12L,LiYun13L,CongjunWu13,CongjunWu11}. (a) For $\l>1$, the superfluid state is an equal superposition of the $k=\pm\a$ states and the order parameter has the form $(\lr{a_{x\ua}},\lr{a_{x\da}}) \propto (\cos\a x,-\sin\a x)$, which corresponds to spin rotating around the $\hat{y}$ axis with wave vector $2\a$. This is the $xy$-chiral phase after rotating around $\hat{x}$ by $\frac{\pi}{2}$. In this limit, the chiral correlation $A^z(i,j)$ is independent of the separation between $i$ and $j$ and is a constant of magnitude $\sin(2\a)/16$, where the factor $16$ comes from the normalization of the spin $s=\frac{1}{2}$. (b) For $\l<1$, however, the weak-coupling superfluid breaks $Z_2$ symmetry by selecting one of the single-particle minima. In this case, the superfluid state is a ferromagnet along the $\hat{z}$ direction (after rotating around $\hat{x}$ by $\pi/2$), consistent with the magnetic order in the Mott phase and confirmed with DMRG calculations.

\begin{figure}[tbp]
\includegraphics[width=0.51 \textwidth] {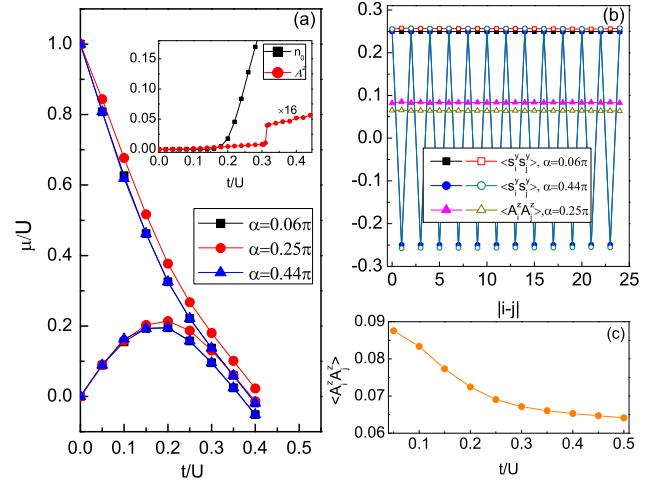}
\caption{(Color online) (a) Phase diagrams for Mott insulator-superfluid
transitions at unit filling with $\alpha=0.06\pi, \l=1.5$ ($\hat{y}$-FM), $\a=0.44\pi,\l=1.5$ ($\hat{y}$-AFM), and $\a=0.25\pi, \l=1.5$ ($xy$ chiral). Different magnetic structures only slightly modify the Mott-superfluid transition boundary. The inset shows the magnetic phase transition between $\hat{y}$-FM and $xy$-chiral phases within the superfluid regime. The inset shows the magnetic phase transition between $\hat{y}$-FM and $xy$-chiral phases within the superfluid regime, $\a=0.08\pi, \l=1.5$. (b) Comparison of various correlation functions in the superfluid ($t/U=0.5$) and Mott insulating phases ($t/U=0.1$). We note the smooth evolution of the magnetic structure across the phase transition. (c) The saturated value of the chiral correlation function $\mathcal{A}^z$ decreases with increasing value of $t/U$, and in this particular case, saturates to a value $\frac{1}{16}=0.625$ for $\a=\frac{\pi}{4}$. The system length for the DMRG calculation is $L=32$.}
\label{fig4}
\end{figure}

With increasing $U$, however, other magnetic phases ($\hat{y}$-FM, $\hat{y}$-AFM for $\l>1$, and PM for $\l<1$) emerge in the superfluid phase, which connect smoothly to those in the Mott insulating phase. We first determine the phase boundaries of the Mott-superfluid transition. Let us define two chemical potentials $\mu_{+}=E(N+1)-E(N)$ and $\mu_{-}=E(N)-E(N-1)$ for a fixed system size. When the chemical potential $\mu$ equals $\mu_{+}$ or $\mu_{-}$, where the phase transition occurs, the quasiparticle or quasihole excitation energy is zero. In Fig.~\ref{fig4}(a), we show the phase diagram in the $\mu - t/U$ plane for different values of $\a$ with $\l=1.5$. In the $\hat{y}$-FM phase ($\a=0.06\pi$) and $\hat{y}$-AFM phase ($\a=0.44\pi$), the phase boundaries are essentially identical within our numerical precision. For the $xy$-chiral state ($\a=0.25\pi$), however, the phase boundary is slightly, but consistently pushed to higher chemical potential. To compare explicitly the magnetic structures in the Mott and superfluid regimes, we plot in Fig.~\ref{fig4}(b) various correlation functions for $t/U=0.1$ (Mott regime, solid markers) and $t/U=0.5$ (superfluid regime, hollow markers) for various values of $\a$ with $\lambda=1.5$. It can be readily observed that apart from quantitative changes in the magnitudes of the correlation functions, both $\mathcal{A}^\g(i,j)$ and $\mathcal{S}^\g(i,j)$ have the same spatial dependence in the Mott and superfluid phases. In Fig.~\ref{fig4}(c), the asymptotic value of the chiral correlation function $\mathcal{A}^z$ is plotted as a function of $t/U$ for $\a=\frac{1}{4}\pi$. We note that the chiral correlation decays as one increases $t/U$, and saturates towards the weak-coupling value $1/16$, as we determined above.

To determine the magnetic phase transitions within the superfluid phase for unit filling, we define $n_0$ to be the maximal amplitude of the one-body density matrix $\langle a_{i\s}^\dag a_{j\s'}\rangle$ that decays algebraically and calculate the chiral order parameter $\mathcal{A}^z$. Here we choose $\a=0.08\pi,\l=1.5$, closer to the $\hat{y}$-FM and $xy$-chiral phase boundary. In the inset of Fig.~\ref{fig4}(a), we observe, with increasing $t/U$, first a second order transition point $t/U\approx 0.15$ from the Mott to the superfluid phase and later, at $t/U\approx 0.31$, a first order phase transition to the chiral superfluid phase, consistent with the weak-coupling phase $U\to 0$.

{\em Conclusions}. We have shown that spin-orbit coupling substantially modifies the Mott insulating phase, resulting in a host of magnetic phases (paramagnetic, ferromagnetic, antiferromagnetic and chiral magnetic). We have discussed how these phases evolve smoothly into the superfluid and, in particular, have shown the existence of a chiral superfluid state in which the phase of the order parameter rotates along the one-dimensional chain. Phase transitions between different magnetically ordered superfluid states are discussed, and we have shown that the transition between the $\hat{y}$-FM superfluid (with no weak-coupling analog) and the $xy$-chiral superfluid is first order.

To detect the magnetic structures in either the Mott or superfluid phases, one can make use of the optical Bragg scattering technique~\cite{Corcovilos2010} and {\em in situ} microscopy which can detect lattice-resolved hyperfine states~\cite{Bakr2010,Sherson2010,Weitenberg2011}. The chiral superfluid can also be identified from measurements of the spin-resolved momentum distributions.

{\em Note added}. Recently, we became aware of the similar DMRG calculations of~\cite{JiZeZhao,JiZeZhao14,Piraud,Peotta}. Our results and theirs agree for the parameter regimes where they overlap.

{\em Acknowledgments}. We thank C. Cheng for helpful discussions. The work described in this Rapid Communication was partially supported by a grant from the RGC, Grant No. HKU-709313P, and a startup grant from University of Hong Kong. W.S.C. acknowledges funding from Grant No. NSF DMR-1309461.


\begin{thebibliography}{99}
\bibitem{com1} For DMRG calculation of the effective spin model, we use open boundary condition and error truncation of the reduced density matrix up to $10^{-6} - 10^{-12}$. We utilized $4$ to $12$ sweeps to reach the convergence of the seventh digit for the ground energy per site. For computation with the original Hubbard model, we truncate the dimension for local Hilbert space to $16$  and error truncation for the reduced density matrix to $10^{-6}$, with  four to eight sweeps to achieve precision at the fifth digit per site.
\end{thebibliography}
\end{document}